\documentclass[a4paper,12pt]{article}
\usepackage{ulem}
\usepackage{ascmac}
\usepackage{amsthm}
\usepackage{amsmath,amssymb}
\usepackage{type1cm}
\usepackage{comment}
\usepackage{algorithmic}
\usepackage{algorithm}
\theoremstyle{plain}
\newtheorem{theorem}{Theorem}[section]
\newtheorem{lemma}{Lemma}[section]
\newtheorem{proposition}{Proposition}[section]

\newtheorem{claim}{Claim}[section]
\newtheorem{definition}{Definition}[section]

\theoremstyle{definition}

\numberwithin{equation}{section}
\newcommand{\leng}[1]{|#1|}
\newcommand{\msc}[1]{\mbox{{\sc #1}}}
\newcommand{\BQED}{\hfill \hbox{\rule{8pt}{8pt}}}

\title{{\bf {\Large Competitive Analysis of Online Facility Assignment 
for General Layout of Servers on a Line}}}
\author{{\sc Tsubasa Harada}%
\thanks{Department of Mathematical and Computing Science, 
Tokyo Institute of Technology, 2-12-1 Ookayama, Meguro-ku, 
Tokyo 152-8550, Japan. 
{\sf harada.t.ak@m.titech.ac.jp}}\and 
{\sc Toshiya Itoh}%
\thanks{Department of Mathematical and Computing Science, 
Tokyo Institute of Technology, 2-12-1 Ookayama, Meguro-ku, 
Tokyo 152-8550, Japan.  {\sf titoh@c.titech.ac.jp}}}
\date{}
\begin{document}
\maketitle
\noindent {\sf Abstract:}  
In the online facility assignment on a line ${\rm OFAL}(S,c)$ with
a set $S$ of $k$ servers and a capacity $c:S\to\mathbb{N}$, 
each server $s\in S$ with a capacity $c(s)$
is placed on a line,
and a request arrives on a line one-by-one.
The task of an online algorithm is to 
irrevocably match a current request with one of the servers with vacancies 
before the next request arrives.
An algorithm can match up to $c(s)$ requests to a server $s\in S$.

In this paper,
we propose a new online algorithm PTCP (Policy Transition at Critical Point)
for OFAL($S,c$) and
show that PTCP is $(2\alpha(S)+1)$-competitive, where
$\alpha(S)$ is informally the ratio of the diameter of $S$
to the maximum distance between two adjacent servers in $S$.
Depending on the layout of servers,
$\alpha(S)$ ranges from constant (independent of $k$) to $k-1$.
Among all of known algorithms for OFAL($S,c$),
this upper bound on the competitive ratio is the best
when
$\alpha(S)$ is small.

We also show that the competitive ratio of any
MPFS (Most Preferred Free Servers) algorithm \cite{HIM2023}
is at least $2\alpha(S)+1$.
For OFAL($S,c$),
recall that MPFS
is a class of algorithms
whose competitive ratio does not depend on a capacity $c$ and
it includes the natural greedy algorithm and PTCP, etc. 
Thus, this implies that
PTCP is the best for OFAL($S,c$) in the class MPFS.
\section{Introduction} \label{sec-introduction}

The \textit{online facility assignment} (OFA)
or \textit{online transportation} problem
was introduced by
Kalyanasundaram and Pruhs \cite{KalP1995}.
In this problem,
an online algorithm is given a set $S$
of $k$ servers and a capacity $c:S\to\mathbb{N}$,
and receives $n$ requests one-by-one in an online fashion.
The task of an online algorithm 
is to match each request immediately
with one of the $k$ servers.
Note that the number of requests is at most
the sum of each server's capacity, i.e.,
$n\leq\sum_{s\in S}c(s)$.
The maximum number of requests that can be
matched with a server $s\in S$ is $c(s)$, and
the assignment cannot be changed later
once it has been decided.
The cost of matching a request with a server is
determined by the distance between them.
The goal of the problem is to
minimize the sum of the costs of matching $n$ requests.
When the underlying metric space is restricted to be a line,
we refer to such a variant of OFA as OFA on a line (denoted by OFAL). 

This problem OFA has many applications.
Consider a car-sharing service where
there are $k$ car stations and each station $s$
has $c(s)$ available cars.
This service must assign users arriving one after another
to car stations immediately.
It is desirable that all users can use a nearby car station
as many as possible.
OFAL also can be viewed as the following real-world problem:
Consider a rental shop that must rent skis with appropriate length
to skiers.
What kind of algorithm can be used to reduce the gap
between the length of the appropriate skis and the actual rented skis?
In this case, each server $s$ is one type of skis, and
its capacity $c(s)$ is the number of $s$ available for rent.

Ahmed et al. \cite{ARK2020} dealt with classical competitive analysis for OFAL
under the assumption that the servers are evenly spaced.
We refer to the setting as OFAL$_{eq}$.
Ahmed et al. \cite{ARK2020} showed (with rough proofs) that
the natural greedy algorithm
matching a request with its closest available server is $4k$-competitive 
and the {\it Optimal-fill\/} algorithm is $k$-competitive for any $k > 2$.
On the other hand, Itoh et al. \cite{IMS2020,IMS2021} analyzed the competitive 
ratio for OFAL$_{eq}$ with small $k\geq 2$.
They showed that (i) for $k=2$, the greedy algorithm is 3-competitive
and best possible, and 
(ii) for $k=3$, $4 $, and $5$, the competitive ratio of any algorithm is at least 
$1+\sqrt{6}>3.449$, $\frac{4+\sqrt{73}}{3}>4.181$, and 
$\frac{13}{3}>4.333$, respectively.

For OFA, Harada et al. \cite{HIM2023} introduced a class of algorithms
called MPFS (Most Preferred Free Servers) as
a natural generalization of the greedy algorithm and
showed that the competitive ratio of any MPFS algorithm
does not depend on a capacity $c$.
This is referred to as the capacity-insensitive property.
In addition,
they determine the exact competitive ratio 
of the greedy algorithm for OFAL$_{eq}$ to be $4k-5$ $(k\geq2)$
by using the properties of MPFS algorithms.
Moreover, they present an MPFS algorithm
IDAS (Interior Division for Adjacent Servers) for OFAL
and showed that the competitive ratio of IDAS for OFAL$_{eq}$
is $2k-1$ and best possible among all MPFS algorithms for OFAL$_{eq}$.

%
\subsection{Our Contributions} \label{subsec-contribution}
%
In this paper, we present a new MPFS algorithm
PTCP (Policy Transition at Critical Point) for OFAL
with a set $S$ of $k$ servers and a capacity $c:S\to\mathbb{N}$
and show that the competitive ratio of PTCP is exactly $2\alpha(S)+1$
and best possible among all MPFS algorithms for OFAL.
As we have mentioned,
IDAS \cite{HIM2023} is best possible only for \uline{OFAL$_{eq}$}.
Informally, $\alpha(S)$ is the ratio of
the diameter of $S$ to the maximum distance
between two adjacent servers in $S$
(see (\ref{ls-alphas}) for details).
We emphasize that PTCP has
the capacity-insensitive property.
Note that $\alpha(S)$ is a constant when
the distances between adjacent servers increase exponentially,
and becomes up to $k-1$ when $k$ servers are evenly placed.

We already have three upper bounds on the competitive ratio for OFAL
with a set $S$ of $k$ servers and a capacity $c$.
These upper bounds can be optimal
according to the specific layout of servers.
\begin{itemize}
\item[(1)] $2^k-1$ (achieved by the greedy algorithm \cite{KalP1998, HIM2023}),
\item[(2)] $2U(S)+1$ (achieved by IDAS \cite{HIM2023}) and
\item[(3)] $O(\log c(S))$
(achieved by Robust-Matching \cite{R2016, R2018}),
\end{itemize}
where $U(S)$ is the aspect ratio of $S$, i.e.,
the ratio of the diameter of $S$ to the minimum distance
between two adjacent servers in $S$ and $c(S):=\sum_{s\in S}c(s)$.

Let us compare the upper bound of PTCP to the above three upper bounds.
For the upper bound (1),
it follows that $2\alpha(S)+1\leq 2k-1 < 2^k-1$ if $k\geq3$ and
$2\alpha(S)+1 = 2^2-1 = 3$ if $k=2$.
Then, our new algorithm PTCP performs better than the greedy algorithm
when $k\geq3$ and performs as well as the greedy algorithm when $k=2$.
For the upper bound (2), we always have $\alpha(S)\leq k-1\leq U(S)$
and the equalities hold if and only if $k$ servers are evenly placed.
Hence, PTCP is better than IDAS except for OFAL$_{eq}$
and performs as well as IDAS for OFAL$_{eq}$.
For the upper bound (3),
there are cases when PTCP performs worse than
Robust-Matching, for example,
the case where the servers are evenly placed and
the capacity of each server is 1.
In this case, we have $O(\log c(S)) = O(\log k) \leq k-1=\alpha(S)$.
However, two cases can be given
where the performance of PTCP is better than that of Robust-Matching.
The first case is when the capacity of each server is very large,
especially when $c(S)=\Omega(2^{\alpha(S)})$, and
the second case is when $\alpha(S)$ is small,
especially when $\alpha(S)=o(\log k)$.

Furthermore, we observe that PTCP is advantageous
against the existing algorithms.
In fact, we show that there exists a layout of servers
such that PTCP performs very well but 
the greedy algorithm performs very poorly (in Theorem \ref{thm-to-greedy}),
and there exists another layout of servers such that
PTCP performs well but
the permutation algorithm \cite{KalP1993, ARK2020}
performs poorly (in Theorem \ref{thm-to-perm}). 

%
\subsection{Related Work} \label{subsec-related}
%
Kalyanasundaram and Pruhs \cite{KalP1995} studied
OFA under the weakened adversary model
where the adversary has only half as many capacities of each server
as the online algorithm and the length of a request sequence is
at most $c(S)/2$ where $c(S)=\sum_{s\in S}c(s)$.
They showed that the greedy algorithm is
$\Theta(\min (k, \log c(S)))$-competitive and
present an $O(1)$-competitive algorithm under this assumption.
Chung et al. \cite{CKP2008} also studied OFA under
another weakened adversary where
the adversary has one less capacity of each server
against the online algorithm.
Under this model,
they presented an $O(\log k)$-competitive deterministic
algorithm on an $\alpha$-HST \cite{CKP2008} metric
where $\alpha=\Omega(\log k)$
and an $O(\log^3 k)$-competitive randomized algorithm
on a general metric.

Ahmed et al. \cite{ARK2020} also considered OFA
on a unweighted graph $G$, and showed (with rough proofs) 
the competitive ratios of the greedy algorithm and 
the Optimal-fill algorithm are $2\leng{E(G)}$ and $\frac{\leng{E(G)}k}{r}$, respectively, 
where $r$ is the radius of $G$.
Muttakee et al. \cite{MAR2020} 
derived (with rough proofs) the competitive ratios of
the greedy algorithm and the Optimal-fill algorithm 
for grid graphs and the competitive ratio of the Optimal-fill algorithm 
for arbitrary graphs.

The special cases of OFA and OFAL, where the capacity of each server is 1,
have been known as the \textit{online metric matching} problem (OMM)
and the \textit{online matching problem on a line} (OML)
respectively.
For OMM, 
Kalyanasundaram and Pruhs \cite{KalP1993} and 
Khuller et al. \cite{KMV1994} presented a deterministic online algorithm, which is 
called {\it Permutation} \cite{KalP1993}, 
and showed that it is $(2k-1)$-competitive and best possible.
In addition, Kalyanasundaram and Pruhs \cite{KalP1993}
also determined the exact competitive ratio of the greedy algorithm
to be $2^k-1$.
The best randomized algorithm for OMM so far \cite{BBGN2007} is
$O(\log^2k)$-competitive and the best lower bound
on the competitive ratio \cite{BBGN2007} is $\Omega(\log k)$.
For doubling metrics, an $O(\log k)$-competitive algorithm \cite{GL2012}
is known.
For OML, there have been many active studies
\cite{AFT2018,ABNPS2014,GL2012,NR2017,R2016, R2018}  
and the best upper bound on the competitive ratio \cite{NR2017, R2018} 
is $O(\log k)$, which is achieved by the deterministic algorithm called
Robust-Matching \cite{R2016}.
The best lower bound on the competitive ratio \cite{PS2021} is 
$\Omega(\sqrt{\log k})$.

There have been extensive studies for the online metric matching
with {\it recourse\/} \cite{GKS2020,MN2020}.
In this problem, an online algorithm is allowed to change
a small number of previous assignments upon arrival of a new request.
For general metrics,
Gupta et al. \cite{GKS2020} showed a deterministic
$O(\log k)$-competitive algorithm with $O(\log k)$-amortized recourse.
For line metrics, they also presented a deterministic
$3$-competitive algorithm with $O(\log k)$-recourse.

Another version of the online bipartite matching problem was initiated by 
Karp et al. \cite{KVV1990}. Since it has application to ad auction, 
several variants of the problem have been  extensively studied (see, e.g., 
\cite{M2012} for a survey). 

%
\section{Preliminaries} \label{sec-preliminary}
%
\subsection{Online Facility Assignment Problem} \label{subsec-ofa}
%
Let $(X,d)$ be a metric space, where $X$ is a (possibly infinite) set of points and 
$d: X \times X \to \mathbb{R}$ is a distance function. 
We use  $S=\{s_{1},\ldots,s_{k}\}$ to denote the set of $k$ servers and use 
$\sigma=r_{1}\cdots r_{n}$ to denote 
a request sequence. For each $1 \leq j \leq k$,  a server 
$s_{j}$ is characterized by the position $p(s_{j}) \in X$ and 
$s_{j}$ has capacity 
$c(s_j) \in \mathbb{N}$, i.e., $s_j$ can be matched with at most $c(s_j)$ requests. 
We assume that $n \leq c(s_1)+\cdots+c(s_k)$. 
For each $1 \leq i \leq n$, 
a request $r_{i}$ is also characterized by the position $p(r_{i}) \in X$. 

The set $S$ is given to an online algorithm in advance, while requests are given 
one-by-one from $r_{1}$ to $r_{n}$. At any time of the execution of an algorithm, 
a server is called {\it free\/} if the number of requests matched with it is less 
than its capacity, and {\it full\/} otherwise. When a request $r_{i}$ is revealed, 
an online algorithm must match $r_{i}$ with one of free servers. If 
$r_{i}$ is matched with the server $s_{j}$, the pair $(r_{i},s_{j})$
is added to the current matching and the cost 
$\msc{cost}(r_{i},s_{j})=d(p(r_{i}),p(s_{j}))$ is incurred for this pair.
The cost of the matching is the sum of the costs of all the pairs contained in it.
The goal of online algorithms is to minimize the cost of the final matching. 
We refer to such a problem as the {\it online facility assignment\/} problem 
with servers $S$ and a capacity $c:S\to \mathbb{N}$
and denote it by ${\rm OFA}(S,c)$. 
For the case that $c(s_1)=\cdots=c(s_{k})=\ell\geq 1$, it is immediate 
that $n \leq k\ell$ and 
we simply use ${\rm OFA}(S,\ell)$ to denote the online facility assignment 
problem with servers $S$ (of uniform capacity $\ell$). 
%
\subsection{Online Facility Assignment Problem on a Line} \label{subsec-ofal}
%
By setting $X=\mathbb{R}$, we can regard 
the online facility assignment problem with servers $S$
as the online facility assignment problem {\it on a line\/} with servers $S$, and 
we denote such a problem by 
${\rm OFAL}(S,c)$ for a general capacity $c:S\to\mathbb{N}$ and 
${\rm OFAL}(S,\ell)$ for a uniform capacity. 
In this case, it is immediate that 
$p(s_{j}) \in \mathbb{R}$ for each $1 \leq j \leq k$ 
and $p(r_{i}) \in \mathbb{R}$ for each $1 \leq i \leq n$. 
Without loss of generality, we assume that $p(s_{1}) < \cdots < p(s_{k})$.

To precisely describe the upper bound of the competitive ratio, 
we introduce the following notation:
for any $T=\{t_1,\ldots ,t_m\}\subseteq S$
where $p(t_1)<\cdots <p(t_m)$, let
\begin{equation}
\label{ls-alphas}
L(T) := \frac{p(t_m)-p(t_1)}{\max_{u}(p(t_{u+1})-p(t_{u}))} \text{ and }
\alpha(S) := \max_{T \subseteq S} L(T).
\end{equation}
For convenience, let $L(S)=0$ and $\alpha(S)=0$ if $|S|\leq1$.

In the rest of the paper, we will abuse the notations 
$r_{i} \in \mathbb{R}$ and $s_{j} \in \mathbb{R}$ for ${\rm OFAL}(S,c)$ 
instead of $p(r_{i}) \in \mathbb{R}$ and $p(s_{j}) \in \mathbb{R}$, respectively, 
when those are clear from the context. 
%
\subsection{Notations and Terminologies} 
\label{subsec-notation-terminology}
%
For a request sequence $\sigma$,
let $|\sigma|$ be the number of requests in $\sigma$.
For an (online/offline) algorithm $\mathcal{A}$ for ${\rm OFA}(S,c)$ 
and a request sequence 
$\sigma=r_{1}\cdots r_{n}$,  
we use $s_{\mathcal{A}}(r_{i};\sigma)$ to denote 
the server with which $\mathcal{A}$ matches $r_{i}$ for each $1 \leq i \leq n$ when 
$\mathcal{A}$ processes $\sigma$. 
Let $\mathcal{A}(\sigma|S)$ be the total cost incurred
when $\mathcal{A}$ processes $\sigma$.
We use {\rm Opt} to denote the optimal {\it offline\/} algorithm, i.e., 
{\rm Opt} knows the entire sequence $\sigma=r_{1}\cdots r_{n}$ 
in advance and {\it minimizes\/} the total cost incurred by {\rm Opt} to 
match each request $r_{i}$ with the server $s_{\rm Opt}(r_{i};\sigma)$. 
Let $F_i(\mathcal{A})$ be the set of all free servers just after
$\mathcal{A}$ matches $r_i$.

To evaluate the performance of an online algorithm $\mathcal{A}$, 
we use the (strict) competitive ratio. 
We say that $\mathcal{A}$ is $\alpha$-competitive if 
$\mathcal{A}(\sigma|S) \leq \alpha \cdot \mathrm{Opt}(\sigma|S)$ for 
any request sequence $\sigma$. 
The competitive ratio ${\cal R}(\mathcal{A})$ of $\mathcal{A}$ is defined 
to be the infimum of $\alpha\geq 1$ such that $\mathcal{A}$ is $\alpha$-competitive, i.e., 
${\cal R}(\mathcal{A}) = \inf \{\alpha \geq 1: \mbox{$\mathcal{A}$ is $\alpha$-competitive}\}$.

%
\subsection{Technical Lemmas} \label{subsec-technical}
%
In this subsection, we introduce some important notions about $\mathrm{OFAL}(S,c)$:
MPFS (Most Preferred Free Servers) algorithm \cite{HIM2023},
surrounding-oriented algorithm \cite{AFT2018,IMS2020,IMS2021}, 
and faithful algorithm \cite{HIM2023}.
In this paper, we mainly deal with surrounding-oriented and faithful MPFS algorithms.
To begin with, we state the definition of an MPFS algorithm and its significant property.
\begin{definition} \label{def-mpfs}
Let $\mathcal{A}$ be an online algorithm for ${\rm OFA}(S,c)$. We say that 
$\mathcal{A}$ is an {\sf MPFS (most preferred free servers)} algorithm if it is 
specified by the following conditions:
Let $\sigma=r_1\ldots r_n$ be a request sequence.
\begin{enumerate}
\item For each $1 \leq i \leq n$, 
the priority of all servers for $r_{i}$ is determined by only $p(r_{i})$, \vspace*{-0.25cm}
\item $\mathcal{A}$ matches $r_{i}$ with a server with the highest priority 
among free servers.
\end{enumerate}
\end{definition}

\noindent
Let ${\cal MPFS}$ be the class of MPFS algorithms.
For each MPFS algorithm $\mathcal{A}$, the following strong theorem
\cite{HIM2023} is known.

\begin{theorem} \label{thm-mpfs}
Let $\mathcal{A}\in \mathcal{MPFS}$ and suppose that 
$\mathcal{A}$ is $\alpha$-competitive for $\mathrm{OFA}(S,1)$.
Then, for any capacity $c:S\to\mathbb{N}$, 
$\mathcal{A}$ is also $\alpha$-competitive for $\mathrm{OFA}(S,c)$.
\end{theorem} 
By this theorem, it turns out that there is no need to specify 
the capacity of each server 
in evaluating the competitive ratio of an MPFS algorithm.
%

\begin{definition} \label{def-surround}
Given a request $r$ for ${\rm OFAL}(S,c)$, 
the {\sf surrounding servers} for $r$ are $s^{L}$ and $s^{R}$, where 
$s^{L}$ is 
the closest free server to the left of $r$ (if any) and $s^{R}$ is 
the closest free server 
to the right of $r$ (if any). If $r=s$ for some $s \in S$ and $s$ is free, 
then the surrounding 
server of $r$ is only the server $s$. 
\end{definition}

Next, we present the notion of surrounding-oriented algorithms \cite{AFT2018,IMS2020,IMS2021} for $\mathrm{OFAL}(S,c)$.
\begin{definition} \label{def-surround-oriented}
Let $\mathcal{A}$ be an online algorithm for ${\rm OFAL}(S,c)$. We say that 
$\mathcal{A}$ is {\sf surrounding-oriented} for a request sequence 
$\sigma$ if it matches every request $r$ of $\sigma$ with one of the 
surrounding servers of $r$. 
We say that $\mathcal{A}$ is 
{\sf surrounding-oriented} if it is 
{\sf surrounding-oriented} for every request sequence $\sigma$. 
\end{definition}
\noindent
For surrounding-oriented algorithms, the following useful lemma 
\cite{AFT2018,IMS2020,IMS2021} is known. 
\begin{lemma} \label{lemma-surrounding-alg}
Let $\mathcal{A}$ be an online algorithm for ${\rm OFAL}(S,c)$. Then there exists 
a surroun\-ding-oriented algorithm $\mathcal{A}'$ for ${\rm OFAL}(S,c)$ such that 
$\mathcal{A}'(\sigma|S) \leq \mathcal{A}(\sigma|S)$ for any $\sigma$. 
\end{lemma}
\noindent
According to Lemma \ref{lemma-surrounding-alg}, we assume that 
any algorithm for ${\rm OFAL}(S,c)$ is surrounding-oriented
in the rest of this paper if otherwise stated.
%

\begin{definition} \label{def-close}
Let $\mathcal{A}$ be an online/offline algorithm for ${\rm OFAL}(S,c)$ and 
$\sigma=r_{1}\cdots r_n$ and $\tau=q_{1}\cdots q_n$ be 
request sequences. We say that $\tau$ is 
{\sf closer} than $\sigma$ w.r.t. $\mathcal{A}$ if \vspace*{-0.25cm}
\begin{enumerate}
\item for each $1 \leq i \leq n$, $q_{i}$ is not farther than $r_{i}$ to 
$s_{\mathcal{A}}(r_{i};\sigma)$ with which $\mathcal{A}$ matches $r_{i}$, i.e., 
$r_{i} \geq q_{i} \geq s_{\mathcal{A}}(r_{i};\sigma)$ or 
$r_{i} \leq q_{i} \leq s_{\mathcal{A}}(r_{i};\sigma)$ and \vspace*{-0.25cm}
\item there exists $1 \leq i \leq n$ such that 
$q_{i}$ is closer than $r_{i}$ to  
$s_{\mathcal{A}}(r_{i};\sigma)$ with which {\sc alg} matches $r_{i}$, i.e., 
$r_{i} > q_{i} \geq s_{\mathcal{A}}(r_{i};\sigma)$ or 
$r_{i} < q_{i} \leq s_{\mathcal{A}}(r_{i};\sigma)$. 
\end{enumerate}
\end{definition}

Finally, we introduce the notion of a faithful algorithm and its useful property.
\begin{definition} \label{def-faithful}
Let $\mathcal{A}$ be an online/offline algorithm for ${\rm OFAL}(S,c)$. 
For any request sequence $\sigma=r_{1}\cdots r_{n}$ and 
any request sequence $\tau=q_{1}\cdots q_{n}$ that is closer than 
$\sigma$ w.r.t. $\mathcal{A}$, we say that 
$\mathcal{A}$ is {\sf faithful} if $s_{\mathcal{A}}(r_{i};\sigma) = s_{\mathcal{A}}(q_{i};\tau)$ 
for each $1 \leq i \leq n$. 
\end{definition}
\begin{definition} \label{def-opposite}
Let $\mathcal{A}$ be an online algorithm for ${\rm OFAL}(S,c)$ 
and $\sigma=r_{1}\cdots r_{n}$ be a request sequence. We say that 
$\sigma$ is {\sf opposite} w.r.t. $\mathcal{A}$ if 
for each $1 \leq i \leq n$, 
\[
r_{i} \in [s_{\mathcal{A}}(r_{i};\sigma),s_{\rm Opt}(r_{i};\sigma)] ~\vee~
r_{i} \in [s_{\rm Opt}(r_{i};\sigma),s_{\mathcal{A}}(r_{i};\sigma)]. 
\] 
\end{definition}
\noindent The following lemma \cite{HIM2023} holds for an opposite request sequence w.r.t. 
$\mathcal{A}$ for ${\rm OFAL}(S,c)$. 

\begin{lemma} \label{lemma-opposite}
Let $\mathcal{A}$ be a faithful online algorithm for ${\rm OFAL}(S,c)$.  
Then, for any request sequence $\sigma$, there exists an opposite 
$\tau$ w.r.t. $\mathcal{A}$
such that $Rate(\sigma) \leq Rate(\tau)$, where 
\[
Rate(\sigma) = \left\{
\begin{array}{cl}
\frac{\mathcal{A}(\sigma|S)}{\mathrm{Opt}(\sigma|S)} & \mbox{\rm if }
\mathrm{Opt}(\sigma|S)>0,\\
\infty & \mbox{\rm if }\mathrm{Opt}(\sigma|S)=0, \mathcal{A}(\sigma|S)>0,\\
1 & \mbox{\rm if }\mathrm{Opt}(\sigma|S)=\mathcal{A}(\sigma|S)=0.
\end{array} \right.  
\]
\end{lemma}
By the above lemma, it is sufficient to analyze only opposite request sequences
in order to upper bound the competitive ratio of a faithful algorithm.
%
\section{``Hybrid'' Algorithm} \label{sec-hybrid}
%
In this section, we mention the properties of the ``hybrid'' algorithm.
This idea was used in \cite{GL2012} first.
Let $\mathcal{A} \in \mathcal{MPFS}$.
For an integer $i \geq 1$ and a server $s \in F_{i-1}({\cal A})$,
the algorithm ${\cal H}_{i,s}^{\cal A}$ matches the requests
$r_1,\ldots ,r_{i-1}$ with
the same servers as ${\cal A}$, $r_i$ with $s$,
and $r_{i+1}, \ldots ,r_k$ with some servers according to ${\cal A}$.
We call ${\cal H}_{i,s}^{\cal A}$ a hybrid algorithm of ${\cal A}$.
If $s=s_{\cal A}(r_i;\sigma)$, then ${\cal A}$ and ${\cal H}_{i,s}^{\cal A}$ are
completely the same.
Then, in the rest of the paper, we consider the case $s\neq s_{\cal A}(r_i;\sigma)$.
We abbreviate $\mathcal{H}^{\mathcal{A}}_{i,s}$ as $\mathcal{H}_{i,s}$
when $\mathcal{A}$ is clear from the context.
%
\begin{lemma} \label{lem-diff-one}
For $\mathrm{OFA}(S,1)$, let $\mathcal{A}\in\mathcal{MPFS}$
and $\sigma=r_1\ldots r_k$ be an request sequence. 
Suppose $s_{\mathcal{A}}(r_i;\sigma) \neq s$. Then,
there exists some $t^*\geq i$,
$\{a_t\}_{t=i}^{t^*}$, and $\{h_t\}_{t=i}^{t^*}$
such that
\begin{itemize}
\setlength{\leftskip}{0.5cm}
\item[(1)] 
$
F_t(\mathcal{A}) \setminus F_t(\mathcal{H}_{i,s}) = \{a_t\} \text{ and }
F_t(\mathcal{H}_{i,s}) \setminus F_t(\mathcal{A}) = \{h_t\}
$
for each $i \leq t \leq t^*$
\item[(2)]
$F_t(\mathcal{A}) = F_t(\mathcal{H}_{i,s})$ for each $t \geq t^* + 1$.
\end{itemize}

\end{lemma}

\noindent \textbf{Proof:}
We prove the lemma by induction on $t$.
Once $F_{t'}(\mathcal{A}) = F_{t'}(\mathcal{H}_{i,s})$ for some $t'\geq i$,
we have $F_{t}(\mathcal{A}) = F_{t}(\mathcal{H}_{i,s})$ for each $t\geq t'$.
Let $t^*+1$ be the smallest such $t'$.

For the base case $t=i$, it is obvious that
\begin{align*}
F_i(\mathcal{A}) \setminus F_i(\mathcal{H}_{i,s}) = \{s\} \text{ and }
F_i(\mathcal{H}_{i,s}) \setminus F_i(\mathcal{A}) = \{s_{\cal A}(r_i;\sigma)\}.
\end{align*}
Then, we have that $a_i=s$ and $h_i=s_{\cal A}(r_i;\sigma)$.

For the inductive step, we assume that there exists $a_t$ and $h_t$ such that
$
F_t(\mathcal{A}) \setminus F_t(\mathcal{H}_{i,s}) = \{a_t\} \text{ and }
F_t(\mathcal{H}_{i,s}) \setminus F_t(\mathcal{A}) = \{h_t\}.
$
Among the servers in $F_t(\mathcal{A})\cup F_t(\mathcal{H}_{i,s})$,
let $s^{(1)}$ (resp. $s^{(2)}$) be the one
with the highest (resp. the second highest) priority 
determined by ${\cal A}$ and $r_{t+1}$.
We consider the following four cases:
\begin{itemize}
\setlength{\leftskip}{1cm}
\item[{\bf Case 1:}] $\{s^{(1)},s^{(2)}\}=\{a_t,h_t\}$,
\item[{\bf Case 2:}] $s^{(1)}$ is neither $a_t$ nor $h_t$,
\item[{\bf Case 3:}] $s^{(1)}=a_t$ and $s^{(2)}\in F_t(\mathcal{A})\cap F_t(\mathcal{H}_{i,s})$, and
\item[{\bf Case 4:}] $s^{(1)}=h_t$ and $s^{(2)}\in F_t(\mathcal{A})\cap F_t(\mathcal{H}_{i,s})$.
\end{itemize}
In Case 1,
$r_{t+1}$ is matched with $a_t$ by $\mathcal{A}$ and
$h_t$ by $\mathcal{H}_{i,s}$. Thus, $F_{t+1}(\mathcal{A}) = F_{t+1}(\mathcal{H}_{i,s})$.
Then, we have that $t=t^*$ and 
$F_{t'}(\mathcal{A}) = F_{t'}(\mathcal{H}_{i,s})$ for any $t'\geq t^*+1$.
We will show that
\[
F_{t+1}(\mathcal{A}) \setminus F_{t+1}(\mathcal{H}_{i,s}) = \{a_{t+1}\} \text{ and }
F_{t+1}(\mathcal{H}_{i,s}) \setminus F_{t+1}(\mathcal{A}) = \{h_{t+1}\}
\]
for some $a_{t+1}$ and $h_{t+1}$
for each of the following cases.

\begin{itemize}

\setlength{\leftskip}{1cm}

\item[\textbf{Case 2:}] $\mathcal{A}$ and $\mathcal{H}_{i,s}$ match $r_{t+1}$
to the same server $s^{(1)}\in F_t(\mathcal{A})\cap F_t(\mathcal{H}_{i,s})$. Then,
\begin{align*}
F_{t+1}(\mathcal{A}) \setminus F_{t+1}(\mathcal{H}_{i,s}) = \{a_t\} \text{ and }
F_{t+1}(\mathcal{H}_{i,s}) \setminus F_{t+1}(\mathcal{A}) = \{h_t\},
\end{align*}
i.e., $a_{t+1}=a_t$ and $h_{t+1}=h_t$.

\item[\textbf{Case 3:}] $r_{t+1}$ is matched with $a_t$ by $\mathcal{A}$ and
$s^{(2)}\in F_t(\mathcal{A})\cap F_t(\mathcal{H}_{i,s})$ by $\mathcal{H}_{i,s}$.
In this case,
\begin{equation*}
\label{lem1-case3}
F_{t+1}(\mathcal{A}) \setminus F_{t+1}(\mathcal{H}_{i,s}) = \{s^{(2)}\} \text{ and }
F_{t+1}(\mathcal{H}_{i,s}) \setminus F_{t+1}(\mathcal{A}) = \{h_t\},
\end{equation*}
i.e., $a_{t+1}=s^{(2)}$ and $h_{t+1}=h_t$.

\item[\textbf{Case 4:}] $r_{t+1}$ is matched with
$s^{(2)}\in F_t(\mathcal{A})\cap F_t(\mathcal{H}_{i,s})$ by $\mathcal{A}$ and
$h_t$ by $\mathcal{H}_{i,s}$.
In this case,
\begin{equation*}
\label{lem1-case4}
F_{t+1}(\mathcal{A}) \setminus F_{t+1}(\mathcal{H}_{i,s}) = \{a_t\} \text{ and }
F_{t+1}(\mathcal{H}_{i,s}) \setminus F_{t+1}(\mathcal{A}) = \{s^{(2)}\},
\end{equation*}
i.e., $a_{t+1}=a_t$ and $h_{t+1}=s^{(2)}$.
\BQED
\end{itemize}
By the proof of Lemma \ref{lem-diff-one}, it is easy to see that
the following proposition on $\{a_t\}_{t=i}^{t^*}$ and $\{h_t\}_{t=i}^{t^*}$ holds.
%
\begin{proposition} \label{prop-basic-atht}
For $\mathrm{OFA}(S,1)$ let $\mathcal{A}\in\mathcal{MPFS}$ 
and $\sigma$ be an request sequence. 
Suppose $s_{\mathcal{A}}(r_i;\sigma) \neq s$. Then, the following conditions hold:
\begin{itemize}
\setlength{\leftskip}{0.5cm}

\item[(P1)] $a_t = a_{t+1}$ or $h_t = h_{t+1}$ for each $i \leq t \leq t^*-1$,

\item[(P2)] If $a_t \neq a_{t+1}$(resp. $h_t \neq h_{t+1}$), then $r_{t+1}$ is matched with $a_t$(resp. $h_{t+1}$) by $\mathcal{A}$
and with $a_{t+1}$(resp. $h_t$) by $\mathcal{H}_{i, s}$ for each $i \leq t \leq t^*-1$, and

\item[(P3)] $r_{t^*+1}$ is matched with $a_{t^*}$ by $\mathcal{A}$ and with $h_{t^*}$ by $\mathcal{H}_{i,s}$.
\end{itemize}

\end{proposition}

The discussion so far holds for general metrics and 
any MPFS algorithm $\mathcal{A}$.
Next, we state an important lemma that holds for a surrounding-oriented MPFS algorithm on a line metric.
%
\begin{lemma} \label{lem-line-atht}
For $\mathrm{OFAL}(S,1)$,
let $\mathcal{A}$ be a surrounding-oriented MPFS algorithm 
and $\sigma$ be a request sequence. 
If $s_{\mathcal{A}}(r_i;\sigma) \neq s$
and there is no free server between $s_{\mathcal{A}}(r_i;\sigma)$ and $s$, then
there is no free server between $a_t$ and $h_t$ for each $i \leq t \leq t^*$, and
either
\begin{itemize}
\setlength{\leftskip}{0.5cm}
\item[(M1)] \qquad\qquad\qquad\qquad$a_{t^*} \leq \cdots \leq a_i < h_i \leq \cdots \leq h_{t^*}$ or
\item[(M2)] \qquad\qquad\qquad\qquad$h_{t^*} \leq \cdots \leq h_i < a_i \leq \cdots \leq a_{t^*}$.
\end{itemize}
\end{lemma}
\noindent \textbf{Proof:} The proof is by induction on $t$.
Recall the four cases in the proof of Lemma \ref{lem-diff-one}.
Note that there is no need to consider Case 1.

For $t=i$, the statement clearly holds by the assumption of the lemma.
Without loss of generality, let $a_i<h_i$.
Assume that there is no free server between $a_t$ and $h_t$ for some 
$i\leq t\leq t^*-1$ and $a_t<h_t$.
For Case 2, $a_t=a_{t+1}$ and $h_t=h_{t+1}$ hold. Then,
there is no free server between $a_{t+1}$ and $h_{t+1}$
and it follows that $a_{t+1}=a_t<h_t=h_{t+1}$.
For Case 3, let $x_t\in F_t({\cal A})\cap F_t({\cal H}_{i,s})$
be the rightmost free server to the left of $a_t$. 
If $r_{t+1}\leq x_t$, then ${\cal A}$ matches $r_{t+1}$ with $a_t$ that is
not surrounding server of $r_{t+1}$.
This contradicts the fact that $\mathcal{A}$ is surrounding-oriented.
If $r_{t+1}\geq h_t$, then
consider the case where only $a_t$ and $h_t$ are free and
a request $r$ occurs at the same position of $r_{t+1}$.
Since $s^{(1)}=a_t$,
$a_t$ has higher priority than $h_t$ for $r$.
Then, in this case,
$\mathcal{A}$ should match $r$ to $a_t$, the free server with the highest priority.
However, by the assumption of $r_{t+1}\geq h_t$,
$a_t$ is not a surrounding server of $r$.
This contradicts the fact that $\mathcal{A}$ is surrounding-oriented.
Thus, we have $x_t< r_{t+1}< h_t$.
Since ${\cal H}_{i,s}$ matches $r_{t+1}$ with a surrounding server of $r_{t+1}$
in $F_t(\mathcal{A})\cap F_t(\mathcal{H}_{i,s})\not\ni h_t$,
the only candidate for $s^{(2)}$ is $x_t$.
Hence, it follows that
$a_{t+1}=x_t<a_t<h_t=h_{t+1}$ and there is no free server between $a_{t+1}$ and $h_{t+1}$.
For Case 4, let $y_t\in F_t({\cal A})\cap F_t({\cal H}_{i,s})$ be
the leftmost free server to the right of $h_t$. We can prove that
$a_{t+1}=a_t<h_t<y_t=h_{t+1}$ and there is no free server between $a_{t+1}$ and $h_{t+1}$
in the same way as Case 3.
\BQED
%
\section{An Optimal MPFS Algorithm for OFAL} \label{sec-def-alg}
%
In this section, we present a new MPFS algorithm
PTCP (Policy Transition at Critical Point $\mathcal{A}^*$) and
show that PTCP is $(2\alpha(S)+1)$-competitive,
where $\alpha(S)$ is given in (\ref{ls-alphas}).
We consider the following properties of MPFS algorithms for
OFAL($S,c$).
\begin{definition} \label{def-c1c2}
Let $\mathcal{A} \in {\cal MPFS}$ for $\mathrm{OFAL}(T,c)$.
We say that $C(\mathcal{A},T)$ holds
if $\mathcal{A}$ satisfies the following conditions: 
\begin{itemize}
\item[(C1)] $\mathcal{A}$ is faithful,

\item[(C2)] $\mathcal{A}$ is surrounding-oriented, and

\item[(C3)] For $\mathrm{OFAL}(T,1)$,
let $\sigma=r_1\ldots r_{|T|}$ be a request sequence.
For $|T| \geq 2$,
consider the hybrid algorithm $\mathcal{H}^{\mathcal{A}}_{i,s}$
where $s\neq s_{\mathcal{A}}(r_i;\sigma)$ is a surrounding server
\footnote{If the number of surrounding servers of $r_i$ is one,
then $s$ is one of the free servers which is just to the left/right of $s_{\mathcal{A}}(r_i;\sigma)$.}
of $r_i$.
Then, $|h_{t^*}-r_i| \leq \alpha(S)|r_i - a_i|$.
\end{itemize}

\end{definition}
\subsection{A new algorithm: Policy Transition at Critical Point} \label{subsec-pre}
Before presenting the algorithm PTCP, we provide several notations.
For a set $S=\{s_1, \ldots ,s_k\}$ of servers
where $s_1 < \cdots < s_k$ and $\max_{u}(s_{u+1}-s_u) = s_{a+1}-s_a$,
let 
\begin{gather}
S_1:=\{s_1, \ldots ,s_a\}, \:
S_2:=\{s_{a+1}, \ldots s_k\}\text{,  and  }
\label{def-s1s2}
x := \frac{\Delta_2+D}{(\Delta_1+D)+(\Delta_2+D)}\cdot D,
\end{gather}
where $\Delta_1:=s_a-s_1$, $\Delta_2:=s_k-s_{a+1}$, and $D:=s_{a+1}-s_a$.
Note that the value of $x$ is determined by the idea
similar to the algorithm IDAS \cite{HIM2023}.

Let $\mathcal{A}^*[S]$ be $\mathcal{A}^*$ for OFAL($S,c$) and
$\mathcal{A}^*[S](r,F)$ be a server with which
$\mathcal{A}^*[S]$ matches $r$
for a set $F\subseteq S$ of free servers.
By using $\mathcal{A}^*[S_1]$ and $\mathcal{A}^*[S_2]$,
we inductively define $\mathcal{A}^*[S](r,F)$ as follows:
\begin{equation}\label{ptcp-def}
\mathcal{A}^*[S](r,F) = \left\{
\begin{array}{ll}
s & \text{ if } F=\{s\}, \\
\mathcal{A}^*[S_1](r,F\cap S_1) & 
\text{ if } (r\leq s_a+x \text{ and } F\cap S_1\neq\emptyset)
\text{ or } (F\cap S_2=\emptyset), \\
\mathcal{A}^*[S_2](r,F\cap S_2) &
\text{ if } (s_a+x<r \text{ and } F\cap S_2\neq\emptyset)
\text{ or } (F\cap S_1=\emptyset).
\end{array}
\right.
\end{equation}
Informally,
$\mathcal{A}^*[S]$ matches a request that occurs
to the left (resp. right) of $s_a+x$
with a free server in $S_1$ (resp. $S_2$)
according to $\mathcal{A}^*[S_1]$ (resp. $\mathcal{A}^*[S_2]$).
The following important lemma holds for $\mathcal{A}^*$ defined above.
\begin{lemma} \label{lem-c1c2}
Let $S$, $S_1$, and $S_2$ be sets of servers defined in (\ref{def-s1s2}).
If $\mathcal{A}^*$ satisfies $C(\mathcal{A}^*,S_1)$ and $C(\mathcal{A}^*,S_2)$,
then $\mathcal{A}^*$ also satisfies $C(\mathcal{A}^*, S)$.
\end{lemma}
\noindent
\textbf{Proof:}
First, we prove that $\mathcal{A}^*[S]$ satisfies (C1).
Consider the situation where $\mathcal{A}^*[S]$ matches a request $r$ with a server $s$
and observe what happens when a request $q$ (located between $r$ and $s$) occurs instead of $r$.
Suppose $s\in S_1$.
There are two possible cases:
(1) $r\leq s_a+x$, or
(2) $r>s_a+x$ and all servers in $S_2$ are full.
For the case (1), $\mathcal{A}^*[S]$ matches $r$ with $s$ according to $\mathcal{A}^*[S_1]$.
Since $\mathcal{A}^*[S_1]$ is faithful, it turns out that $\mathcal{A}^*[S]$ matches $q$ with $s$.
For the case (2), $s$ is the rightmost free server and $s<r$ holds. 
Therefore, $\mathcal{A}^*[S]$ matches $q$ with $s$.
The same discussion can be applied to the case $s \in S_{2}$.
Hence, $\mathcal{A}^*[S]$ is faithful,
i.e., $\mathcal{A}^*[S]$ satisfies (C1).

Next, we prove that $\mathcal{A}^*[S]$ satisfies (C2).
By contradiction,
assume that there exist a request sequence $r_1\ldots r_n$ and a request $r_t$
such that $\mathcal{A}^*[S]$ matches $r_t$ with a server $s$ that is not a surrounding server of $r_t$.
Let $s'$ be a free server between $r_t$ and $s$. 
If $s,s'\in S_1$ (resp. $s,s'\in S_2$),
then $\mathcal{A}^*[S]$ matches $r_t$ with $s$
according to ${\cal A}^*[S_1]$ (resp. ${\cal A}^*[S_2]$).
However, this contradicts the fact that
${\cal A}^*[S_1]$ (resp. ${\cal A}^*[S_2]$) is surrounding-oriented.
If $s\in S_1$ and $s'\in S_2$, then we have $s<s_a+x<s'\leq r_t$. 
Since $s_a +x<r_t$ and there exists a free server $s'\in S_2$, 
$\mathcal{A}^*[S]$ must match $r_t$ with a free server in $S_2$
according to ${\cal A}^*[S_2]$ and
this contradicts the assumption that 
$\mathcal{A}^*[S]$ matches $r_t$ with $s\in S_1$.
The same discussion can be applied to the case where $s\in S_2$ and $s'\in S_1$.
Therefore, $\mathcal{A}^*[S]$ is surrounding-oriented,
i.e., $\mathcal{A}^*[S]$ satisfies (C2).

Finally, we prove that $\mathcal{A}^*[S]$ satisfies (C3).
Fix any request sequence $\sigma=r_1\ldots r_k$ for OFAL($S,1$).
Let $s\neq s_{\mathcal{A}^*[S]}(r_i;\sigma)$ be a surrounding server of $r_i$
and consider the hybrid algorithm $\mathcal{H}_{i,s}$ of $\mathcal{A}^*[S]$.
By definition, it follows that $a_i=s$ and $h_i=s_{\mathcal{A}^*[S]}(r_i;\sigma)$.
We use the following claim.
%
\begin{claim} \label{claim2}
If $a_{i},h_{i} \in S_{1}$ (resp.  $a_{i},h_{i} \in S_{2}$),
then  $a_{t},h_{t} \in S_{1}$ (resp.  $a_{t},h_{t} \in S_{2}$)
for each $i \leq t \leq t^{*}$ where
$t^*$ is defined in Lemma \ref{lem-diff-one}.
\end{claim}
\noindent \textbf{Proof of Claim \ref{claim2}:}
We prove the above claim for the case $a_{i},h_{i} \in S_{1}$.
The claim for the case $a_{i},h_{i} \in S_{2}$ can be shown analogously.
By contradiction, assume that there exists a time $t'\geq i+1$ such that
$a_{t'}\notin S_1$ or $h_{t'}\notin S_1$.
Let $t$ be the smallest such $t'$.
Then, we have that
(1) $a_t, h_t, h_{t+1}\in S_1, a_{t+1}\in S_2$ or
(2) $h_t, a_t, a_{t+1}\in S_1, h_{t+1}\in S_2$.
Note that $a_t,h_t \in S_1$ and $a_{t+1},h_{t+1}\in S_2$ does not hold
for each $i\leq t\leq t^*-1$
by (P1) of Proposition \ref{prop-basic-atht}.
For the case (1),
by (P2) of Proposition \ref{prop-basic-atht},
$\mathcal{A}^*[S]$ matches $r_{t+1}$ with $a_t\in S_1$.
There are two cases: 
(1.a) $r_{t+1}\leq s_a+x$, or
(1.b) $r_{t+1}>s_a+x$ and all servers in $S_2$ are full,
but the case (1.b) contradicts the assumption $a_{t+1}\in S_2$.
Then, we focus on the case (1.a).
$\mathcal{H}_{i,s}$ has at least one free server in $S_1$
since $h_t \in S_1$.
Then, $\mathcal{H}_{i,s}$ must match
$r_{t+1}$ with a server in $S_1$.
However, 
$\mathcal{H}_{i,s}$ matches $r_{t+1}$ with $a_{t+1}\in S_2$
by (P2) of Proposition \ref{prop-basic-atht}
and this is a contradiction.
The same contradiction can be derived for the case (2). 
\BQED
\vskip.5\baselineskip

If $a_{i},h_{i} \in S_{1}$,
then let $\sigma_1$ be a subsequence of $\sigma$
consisting of all requests $r$ that satisfies
$s_{\mathcal{A}^*[S]}(r;\sigma)\in S_1$.
By the definition of $\mathcal{A}^*[S]$,
each request in $\sigma_1$ is matched with a server in $S_1$
according to $\mathcal{A}^*[S_1]$.
Since $a_t,h_t\in S_1$ for each $i\leq t\leq t^*$ by Claim \ref{claim2},
all requests that may affect the changes
in $\{a_t\}_{t=i}^{t^*}$ and $\{h_t\}_{t=i}^{t^*}$
are included in $\sigma_1$.
In addition, the way of changes in $\{a_t\}_{t=i}^{t^*}$ and $\{h_t\}_{t=i}^{t^*}$
depends only on the behavior of $\mathcal{A}^*[S_1]$.
Then, by the assumption that $C(\mathcal{A}^*,S_1)$ holds,
we can see that $\mathcal{A}^*[S]$ satisfies (C3).
Analogously,
it can be shown that $\mathcal{A}^*[S]$ satisfies (C3) for $a_i,h_i\in S_2$.

Hence, the remaining possible cases are
(1) $a_i\in S_1$ and $h_i\in S_2$,
and
(2) $h_i\in S_1$ and $a_i\in S_2$.
For the case (1), by the fact $a_i\leq s_a<s_a+x\leq r_i\leq h_i$,
it follows that
\begin{align*}
\frac{|h_{t^*}-r_i|}{|r_i-a_i|}
&\leq \frac{\Delta_2+D-x}{x}
=\frac{\Delta_1+\Delta_2+D}{D}
=L(S)
\leq \alpha(S).
\end{align*}
For the case (2), by using the fact $h_i\leq r_i \leq s_a+x <s_{a+1}\leq a_i$, we have
\begin{align*}
\frac{|h_{t^*}-r_i|}{|r_i-a_i|}
&\leq \frac{\Delta_1+x}{D-x}
=\frac{\Delta_1+\Delta_2+D}{D}
=L(S)
\leq \alpha(S).
\end{align*}
The second equality and the last inequality is due to
the definition of $L(S)$ and $\alpha(S)$ in (\ref{ls-alphas}).
Therefore, $\mathcal{A}^*[S]$ satisfies (C3).
\BQED
%
%
%
%
%
\subsection{An Upper Bound on the Competitive Ratio of PTCP} \label{subsec-pf-main}
%
The goal of this subsection is to prove the following theorem, which
claims that PTCP is $(2\alpha(S)+1)$-competitive.
\begin{theorem} \label{thm-main}
For $\mathrm{OFAL}(S,c)$,
$\mathcal{A}^*$ defined in (\ref{ptcp-def}) is $(2\alpha(S)+1)$-competitive,
where $c:S\to \mathbb{N}$ is an arbitrary capacity.
\end{theorem}

To prove Theorem \ref{thm-main}, we introduce a simpler algorithm
similar to $\mathcal{A}^*$ and show the important lemma about the algorithm.
Let $S=\{s_1,\ldots ,s_k\}$ be a set of servers where $s_1<\cdots <s_k$, and
$\mathcal{A}\in\mathcal{MPFS}$ be a $(2\alpha(S)+1)$-competitive algorithm
for which $C(\mathcal{A},S)$ holds. 
Let $d$ and $x$ be parameters such that $s_{k+1}=s_k+d$ and $0<x<d$,
and
define a new MPFS algorithm $\mathcal{A}_{d,x}$ for 
$\mathrm{OFAL}(S \cup \{s_{k+1}\},c)$ as follows:

\begin{itemize}
\item[1.]  If $r \leq s_k+x$, then match a new request $r$ with a server in $S$ according to $\mathcal{A}$. When all servers in $S$ are full just before $r$ is revealed, match $r$ with $s_{k+1}$.
\item[2.]  If $r > s_k+x$, then match a new request $r$ with $s_{k+1}$. When $s_{k+1}$ is full just before $r$ is revealed, match $r$ to a server in $S$ according to $\mathcal{A}$.
\end{itemize}
%
\begin{lemma} \label{lem-endleft}
Let $S=\{s_1,\ldots ,s_k\}$ and $\tilde{S}=\{s_1,\ldots ,s_{k+1}\}$
be sets of servers, where $s_1<\cdots <s_k<s_{k+1}$, and
$\mathcal{A}\in \mathcal{MPFS}$ be a $(2\alpha(S)+1)$-competitive algorithm for which 
$C(\mathcal{A},S)$ holds. Then,
for any request sequence $\sigma$ of $\mathrm{OFAL}(\tilde{S},c)$,
\[
\mathcal{A}_{d,x}(\sigma|\tilde{S}) \leq 
\max\left\{ 2\alpha(S)+1,\frac{2d-x}{x},\frac{2\Delta+d+x}{d-x}\right\}\mathrm{Opt}(\sigma|\tilde{S}),
\]
where $\Delta:=s_k-s_1$, $d=s_{k+1}-s_k$, $0<x<d$,
and $c:\tilde{S}\to\mathbb{N}$ is an arbitrary capacity.
\end{lemma}
\noindent \textbf{Proof:}
Since $\mathcal{A}$ is a faithful MPFS algorithm by (C1),
we can show that $\mathcal{A}_{d,x}$ is also a faithful MPFS algorithm
similarly to the proof of Lemma \ref{lem-c1c2}.
Hence, it suffices to consider 
opposite request sequences for $\mathrm{OFAL}(\tilde{S},1)$.
Let $\sigma$ be any opposite request sequence for $\mathrm{OFAL}(\tilde{S},1)$.
Note that each request of $\sigma$ occurred on $[s_1,s_{k+1}]$.

Let $m$ be the number of requests occurred in $(s_k+x, s_{k+1}]$ in $\sigma$.
If $m>2$, then there exists at least one request $r\in(s_k+x, s_{k+1}]$ such that
$s_{\mathcal{A}_{d,x}}(r;\sigma),s_{\mathrm{Opt}}(r;\sigma)\in S$.
Then, we have
\[
\max\{s_{\mathcal{A}_{d,x}}(r;\sigma),s_{\mathrm{Opt}}(r;\sigma)\}<r,
\]
and this contradicts the assumption that $\sigma$ is opposite.
Hence, there are three cases to be considered: $m=0$, $m=1$, and $m=2$.

We first consider the simplest case $m=1$.
Let $r$ be the unique request occurred in $(s_k+x, s_{k+1}]$
and $\sigma-r$ be a request sequence for $\mathrm{OFAL}(S,1)$
obtained by deleting $r$ from $\sigma$.
Obviously, $r$ is matched with $s_{k+1}$ by both $\mathcal{A}_{d,x}$ and ${\rm Opt}$.
By the definition of $\mathcal{A}_{d,x}$, the assignment of requests in $\sigma-r$
by $\mathcal{A}_{d,x}$ is the same as that by $\mathcal{A}$. Then we have
\begin{align*}
\mathcal{A}_{d,x}(\sigma|\tilde{S}) &= \mathcal{A}(\sigma-r|S) + |r-s_{k+1}| \\
&\leq (2\alpha(S)+1)\mathrm{Opt}(\sigma-r|S) + |r-s_{k+1}| \\
&\leq (2\alpha(S)+1)(\mathrm{Opt}(\sigma-r|S) + |r-s_{k+1}|) \\
&= (2\alpha(S)+1)\mathrm{Opt}(\sigma|\tilde{S}).
\end{align*}

We next consider the case $m=2$.
Let $r_{i_1}$ and $r_{i_2}$ be the two requests
occurred in $(s_k+x, s_{k+1}]$
where $i_1<i_2$.
Since $\mathcal{A}_{d,x}$ matches $r_{i_1}$ with $s_{k+1}$ and
$\sigma$ is opposite,
$\mathrm{Opt}$ matches $r_{i_2}$ with $s_{k+1}$.
Let $\sigma'$ be the request sequence for $\mathrm{OFAL}(S,1)$
obtained by deleting $r_{i_1}$ from $\sigma$
and changing the position of $r_{i_2}$ to $s_{k}$.
Then, it follows that
\begin{align*}
\mathcal{A}_{d,x}(\sigma|\tilde{S}) &= \mathcal{A}(\sigma'|S)
+ |r_{i_1}-s_{k+1}| + |r_{i_2}-s_k| \\
&\leq \mathcal{A}(\sigma'|S) + (d-x) + d \\
&\leq (2\alpha(S)+1)\mathrm{Opt}(\sigma'|S) + 2d-x \\
&\leq \max\left\{2\alpha(S)+1, \frac{2d-x}{x}\right\}(\mathrm{Opt}(\sigma'|S) + x) \\
&\leq \max\left\{2\alpha(S)+1, \frac{2d-x}{x}\right\}(\mathrm{Opt}(\sigma'|S) + |r_{i_2}-s_{k+1}| + |r_{i_1}-s_k|) \\
&= \max\left\{2\alpha(S)+1, \frac{2d-x}{x}\right\}\mathrm{Opt}(\sigma|\tilde{S}).
\end{align*}

Finally, we consider the case $m=0$.
Let $r_i$ be the rightmost request in $\sigma$
and $s^*$ be the rightmost server
in $F_{i-1}(\mathcal{A}_{d,x})\cap S$.
Since $\sigma$ is opposite and $s_{\rm Opt}(r_i;\sigma) = s_{k+1}$, we have
$s_{\mathcal{A}_{d,x}}(r_i;\sigma) \neq s_{k+1}$.
We have two cases:
(1) $s_{\mathcal{A}_{d,x}}(r_i;\sigma)=s^*$ and
(2) $s_{\mathcal{A}_{d,x}}(r_i;\sigma)\neq s^*$.

For the case (1),
consider the hybrid algorithm $\mathcal{H}_{i,s_{k+1}}$ of $\mathcal{A}_{d,x}$.
Recall that $\mathcal{H}_{i,s_{k+1}}$ is an algorithm that matches
$r_1,\ldots ,r_{i-1}$ with the same servers as $\mathcal{A}_{d,x}$, $r_i$ with $s_{k+1}$,
and $r_{i+1},\ldots,r_{k+1}$ with some servers according to $\mathcal{A}_{d,x}$.
Note that $h_{i}=s^{*}\leq r_{i}\leq s_{k+1} = a_{i}$.
By Proposition \ref{prop-basic-atht} and the triangle inequality, it follows that
\begin{align*}
|r_{t+1}-&s_{\mathcal{A}_{d,x}}(r_{t+1};\sigma)|-|r_{t+1}-s_{\mathcal{H}_{i,s_{k+1}}}(r_{t+1};\sigma)|\\
&= |r_{t+1}-a_{t}|-|r_{t+1}-a_{t+1}|+|r_{t+1}-h_{t+1}|-|r_{t+1}-h_{t}| \\
&\leq |a_{t}-a_{t+1}| + |h_{t}-h_{t+1}| 
\end{align*}
for each $i\leq t\leq t^*-1$ and
\begin{align*}
|r_{t^*+1}-s_{\mathcal{A}_{d,x}}&(r_{t^*+1};\sigma)|-|r_{t^*+1}-s_{\mathcal{H}_{i,s_{k+1}}}(r_{t^*+1};\sigma)| \\
&=|r_{t^*+1}-a_{t^*}|-|r_{t^*+1}-h_{t^*}| \leq |a_{t^*}-h_{t^*}|.
\end{align*}
Thus, by Lemma \ref{lem-line-atht}, we have
\begin{align*}
&\mathcal{A}_{d,x}(\sigma|\tilde{S}) - \mathcal{H}_{i,s_{k+1}}(\sigma|\tilde{S})
= \sum_{t=i}^{t^*+1} \left( |r_t-s_{\mathcal{A}_{d,x}}(r_t;\sigma)|-|r_t-s_{\mathcal{H}_{i,s_{k+1}}}(r_t;\sigma)| \right) \\
&\hspace{2cm}\leq |r_i-h_i|-|r_i-a_i|+
\sum_{t=i+1}^{t^*} \left(|a_{t-1}-a_t| + |h_{t-1}-h_t| \right) 
+ |a_{t^*}-h_{t^*}| \\
&\hspace{2cm}= |r_i-h_i|-|r_i-a_i|+2\,|a_{t^*}-h_{t^*}|-|a_i-h_i| \\
&\hspace{2cm}= 2\,(|a_{t^*}-h_{t^*}|-|r_i-a_i|) \\
&\hspace{2cm}\leq 2\,\left((\Delta+d) -(d-x)\right) \\
&\hspace{2cm}= \frac{2\,(\Delta+x)}{d-x}(d-x) \leq \frac{2\,(\Delta+x)}{d-x}|r_i-s_{k+1}|.
\end{align*}
Let $\sigma'$ be a request sequence for $\mathrm{OFAL}(S,1)$
obtained by deleting $r_i$ from $\sigma$.
By the definition of $\mathcal{A}_{d,x}$, the following two formulas hold:
\begin{align*}
\mathcal{H}_{i,s_{k+1}}(\sigma|\tilde{S})&=\mathcal{A}(\sigma'|S)+|r_i-s_{k+1}|, \\
\mathrm{Opt}(\sigma|\tilde{S})&=\mathrm{Opt}(\sigma'|S)+|r_i-s_{k+1}|.
\end{align*} 
Therefore, we finally obtain
\begin{align*}
\mathcal{A}_{d,x}(\sigma|\tilde{S})&=
\mathcal{A}_{d,x}(\sigma|\tilde{S})-\mathcal{H}_{i,s_{k+1}}(\sigma|\tilde{S})
+\mathcal{A}(\sigma'|S)+|r_i-s_{k+1}| \\
&\leq \frac{2\,(\Delta+x)}{d-x}|r_i-s_{k+1}|+(2\alpha(S)+1)\mathrm{Opt}(\sigma'|S)
+|r_i-s_{k+1}| \\
&=\frac{2\Delta+d+x}{d-x}|r_i-s_{k+1}|+(2\alpha(S)+1)
\left(\mathrm{Opt}(\sigma|\tilde{S})-|r_i-s_{k+1}|\right) \\
&\leq \max\left\{ 2\alpha(S)+1, \frac{2\Delta+d+x}{d-x}\right\}\mathrm{Opt}(\sigma|\tilde{S}).
\end{align*}

For the case (2),
we have $r_i<s^*$ and the following claim holds.
%
\begin{claim} \label{claim1}
Let $\sigma=r_1\ldots r_{k+1}$ be an opposite request sequence for $\mathrm{OFAL}(\tilde{S},1)$,
$r_i$ be the rightmost request in $\sigma$, 
and $s^*$ be the rightmost server in $F_{i-1}(\mathcal{A}_{d,x})\cap S$.
Define $\sigma^*$ to be the request sequence for $\mathrm{OFAL}(\tilde{S})$
that changes the position of $r_i$ in $\sigma$ to $s^*$.
If $r_i<s^*$, then we have
\[
\mathcal{A}_{d,x}(\sigma|\tilde{S}) - \mathcal{A}_{d,x}(\sigma^*|\tilde{S}) \leq (2\alpha(S)+1)|r_i-s^*|.
\]
\end{claim}
%
\noindent The proof of Claim \ref{claim1} is given in Appendix \ref{app-pf-claim1}.
The discussion so far can be applied to $\sigma^*=r_1^*\cdots r_{k+1}^*$
since $s_{\mathcal{A}_{d,x}}(r^*_i;\sigma^*)= s^*$.
Hence, we have
\[
\mathcal{A}_{d,x}(\sigma^*|\tilde{S}) \leq
\max\left\{ 2\alpha(S)+1, \frac{2\Delta+d+x}{d-x}\right\}\mathrm{Opt}(\sigma^*|\tilde{S})
\]
and $\mathrm{Opt}(\sigma|\tilde{S})=\mathrm{Opt}(\sigma^*|\tilde{S})+|r_i-s^*|$.
Then, we obtain
\begin{align*}
\mathcal{A}_{d,x}(\sigma|\tilde{S})&=
\mathcal{A}_{d,x}(\sigma|\tilde{S}) - \mathcal{A}_{d,x}(\sigma^*|\tilde{S})
+\mathcal{A}_{d,x}(\sigma^*|\tilde{S})\\
&\leq (2\alpha(S)+1)|r_i-s^*|+
\max\left\{ 2\alpha(S)+1, \frac{2\Delta+d+x}{d-x}\right\}\mathrm{Opt}(\sigma^*|\tilde{S})\\
&\leq \max\left\{ 2\alpha(S)+1, \frac{2\Delta+d+x}{d-x}\right\}\left(\mathrm{Opt}(\sigma^*|\tilde{S})+|r_i-s^*|\right)\\
&=\max\left\{ 2\alpha(S)+1, \frac{2\Delta+d+x}{d-x}\right\}\mathrm{Opt}(\sigma|\tilde{S}).
\end{align*}

Thus, for any opposite request sequence $\sigma$, it follows that
\[
\mathcal{A}_{d,x}(\sigma|\tilde{S}) \leq 
\max\left\{ 2\alpha(S)+1,\frac{2d-x}{x},\frac{2\Delta+d+x}{d-x}\right\}\mathrm{Opt}(\sigma|\tilde{S}).
\]
This completes the proof of the lemma.
\BQED
\vskip.5 \baselineskip
Now we are ready to prove the main theorem.
\vskip.5 \baselineskip
\noindent \textbf{Proof of Theorem \ref{thm-main}:}
The proof is by induction on the number $k$ of servers.
For the base case $k=1$, 
$\mathcal{A}^*[\{s\}]$ is a trivial algorithm that
matches every request to the unique server $s$.
Then, $C(\mathcal{A}^*,\{s\})$ holds and 
$\mathcal{A}^*[\{s\}]$ is $2\alpha(\{s\})+1=1$-competitive.

For the inductive step,
assume that $\mathcal{A}^*[T]$ is $(2\alpha(T)+1)$-competitive
and $C(\mathcal{A}^*, T)$ holds
for any set $T$ of servers such that $|T|\leq k-1$.
Let $S=\{s_1,\ldots ,s_k\}$ be a set of $k\geq2$ servers
where $s_1<\cdots <s_k$ and
$a$ be any integer such that $\max_u(s_{u+1}-s_u)=s_{a+1}-s_a$.
Note that
$S_1$, $S_2$, $\Delta_1$, $\Delta_2$ and $x$
are given by (\ref{def-s1s2}).
By the induction hypothesis,
$\mathcal{A}^*[S_1]$ (resp. $\mathcal{A}^*[S_2]$) is
$(2\alpha(S_1)+1)$-competitive (resp. $(2\alpha(S_2)+1)$-competitive)
and $C(\mathcal{A}^*, S_1)$ (resp. $C(\mathcal{A}^*, S_2)$) holds.
Fix a request sequence $\sigma$ arbitrarily and
let $m$ be the number of requests in $\sigma$ that occur in $(-\infty, s_a+x]$.
There are three cases to be considered: $m=a$, $m>a$ and $m<a$.

For the first case $m=a$,
define $\sigma_1$ (resp. $\sigma_2$) to be a subsequence of $\sigma$
consisting of all requests $r$
such that $r \leq s_a+x$ (resp. $s_a+x<r$).
Since $\mathcal{A}^*$ and $\mathrm{Opt}$ match each request
in $\sigma_1$ (resp. $\sigma_2$) with a server in $S_1$ (resp. $S_2$),
we have that
\begin{align*}
\mathcal{A}^*(\sigma|S) 
&= \mathcal{A}^*(\sigma_1|S_1) + \mathcal{A}^*(\sigma_2|S_2)\\
&\leq (2\alpha(S_1)+1)\mathrm{Opt}(\sigma_1|S_1) + 
(2\alpha(S_2)+1)\mathrm{Opt}(\sigma_2|S_2)\\
&\leq \left(2\max\left\{\alpha(S_1),\alpha(S_2)\right\}+1\right)\left(\mathrm{Opt}(\sigma_1|S_1)+\mathrm{Opt}(\sigma_2|S_2)\right)\\
&\leq(2\alpha(S)+1)\mathrm{Opt}(\sigma|S).
\end{align*}

For the second case $m>a$,
define subsequences $\sigma_{1}$ and $\sigma_{2}$ of $\sigma$ as follows: 
$\sigma_1$ consists of all requests $r$ such that $r \leq s_a+x$ and
$\sigma_2$ consists of all requests $r$ that is matched with a server in $S_2$ by $\mathcal{A}^*$.
Let $R$ be a set of requests consisting of all requests 
that belong to both $\sigma_1$ and $\sigma_2$.
Note that $R$ consists of the last $|\sigma_1|-a$ requests in $\sigma_1$.
We use $\sigma'_2$ to denote a request sequence
obtained by moving the position of
each request in $R$ to $s_{a+1}$.
Consider the following operations for $\sigma_1$ and $\sigma'_2$:
Operation (1) $\mathcal{A}^*[S_1]_{D,x}$ serves $\sigma_1$
with servers $\tilde{S_1}=S_1\cup\{s_{a+1}\}$ and a capacity $c_1$
where $c_1(s_j)=1$ for each $1\leq j\leq a$ and $c_1(s_{a+1})=|R|$, and
Operation (2) $\mathcal{A}^*$ serves $\sigma'_2$ with servers $S_2$ and
a capacity $c_2(s)=1$ for each $s\in S_2$. 
By the definition of $\mathcal{A}^*$ and
$\mathcal{A}^*[S_1]_{D,x}$,
we have the following:
\begin{align*}
\mathcal{A}^*(\sigma|S)
&= \mathcal{A}^*[S_1]_{D,x}(\sigma_1|\tilde{S_1})+
\mathcal{A}^*(\sigma'_2|S_2),\\
\mathrm{Opt}(\sigma|S)
&= \mathrm{Opt}(\sigma_1|\tilde{S_1})+\mathrm{Opt}(\sigma'_2|S_2).
\end{align*}
where $\mathcal{A}^*[S_1]_{D,x}(\sigma_1|\tilde{S_1})$
denotes the cost of an algorithm $\mathcal{A}^*[S_1]_{D,x}$
for the operation (1)
and $\mathcal{A}^*(\sigma'_2|S_2)$ denotes the cost of an algorithm
$\mathcal{A}^*$ for the operation (2).
We define
$\mathrm{Opt}(\sigma_1|\tilde{S_1})$ and
$\mathrm{Opt}(\sigma'_2|S_2)$
analogously.
Thus, by Lemma \ref{lem-endleft}, we get
\begin{align}
&\mathcal{A}^*(\sigma|S) 
= \mathcal{A}[S_1]_{D,x}(\sigma_1|\tilde{S_1})
+ \mathcal{A}^*(\sigma'_2|S_2)\nonumber\\
&\hspace{0.5cm}\leq \max\left\{2\alpha(S_1)+1,\frac{2D-x}{x},\frac{2\Delta_1+D+x}{D-x}\right\}\mathrm{Opt}(\sigma_1|\tilde{S_1}) + 
(2\alpha(S_2)+1)\mathrm{Opt}(\sigma'_2|S_2)\nonumber\\
&\hspace{0.5cm}\leq\max\left\{2\alpha(S_1)+1,2\alpha(S_2)+1,\frac{2D-x}{x},\frac{2\Delta_1+D+x}{D-x}\right\}\mathrm{Opt}(\sigma|S).
\label{f1}
\end{align}
By substituting $x=D(\Delta_2+D)/(\Delta_1+\Delta_2+2D)$,
\begin{align*}
\max&\left\{2\alpha(S_1)+1,2\alpha(S_2)+1,\frac{2D-x}{x},\frac{2\Delta_1+D+x}{D-x}\right\}\\
&=\max\left\{2\alpha(S_1)+1,2\alpha(S_2)+1,\frac{2\Delta_1+D+x}{D-x}\right\}\\
&=\max\left\{2\alpha(S_1)+1,2\alpha(S_2)+1,2\frac{\Delta_1+\Delta_2+D}{D}+1\right\}\\
&\leq 2\alpha(S)+1.
\end{align*}
Thus, 
$
\mathcal{A}^*(\sigma|S)\leq(2\alpha(S)+1)\mathrm{Opt}(\sigma|S)
$ for the case $m>a$.

For the last case $m<a$,
we can use the proof for the case $m>a$ by symmetry. 
By replacing $S_1$, $S_2$, $\Delta_1$, and $x$ in (\ref{f1}) with
$S_2$, $S_1$, $\Delta_2$, and $D-x$ respectively, we have
\[
\mathcal{A}^*(\sigma|S)
\leq \max\left\{2\alpha(S_2)+1,2\alpha(S_1)+1,\frac{D+x}{D-x},\frac{2\Delta_2+2D-x}{x}\right\}\mathrm{Opt}(\sigma|S),
\]
and by substituting $x=D(\Delta_2+D)/(\Delta_1+\Delta_2+2D)$, it follows that
\begin{align*}
\max&\left\{2\alpha(S_2)+1,2\alpha(S_1)+1,\frac{D+x}{D-x},\frac{2\Delta_2+2D-x}{x}\right\}\\
&=\max\left\{2\alpha(S_1)+1,2\alpha(S_2)+1,2\frac{\Delta_1+\Delta_2+D}{D}+1\right\}\\
&\leq 2\alpha(S)+1.
\end{align*}
Therefore, we finally obtain
$
\mathcal{A}^*(\sigma|S)\leq(2\alpha(S)+1)\mathrm{Opt}(\sigma|S)
$
for the case $m<a$.
\BQED

\subsection{Comparisons with Other Algorithms}
\label{subsec-comparison}
%
In this subsection, we compare the performance of the PTCP algorithm with
other well-known algorithms for OFAL($S,c$), e.g. the greedy algorithm and
the permutation algorithm \cite{KalP1993, ARK2020}.

\subsubsection{Comparison with the Greedy Algorithm}
\label{subsubsec-to-greedy}
The greedy algorithm (denoted by $\mathcal{G}$) for OFAL($S,c$) is an algorithm 
that matches a newly occurred request to the nearest free server.
For the competitive ratio of $\mathcal{A}^*$ and $\mathcal{G}$,
we have the following theorem, which
implies that there exists a server layout
where $\mathcal{A}^*$ performs very well
while $\mathcal{G}$ performs very poorly.
\begin{theorem} \label{thm-to-greedy}
Define $S=\{s_1,\ldots ,s_k\}$ as follows: $s_1=0$ and $s_i=2^{i-1}$ for $i=2,\ldots ,k$.
For the server layout $S$, $\mathcal{A}^*$ is $5$-competitive
and the competitive ratio of $\mathcal{G}$ is at least $2^k-1$.
\end{theorem}
The proof of Theorem \ref{thm-to-greedy}
can be found in Appendix \ref{pf-to-greedy}.

\subsubsection{Comparison with the Permutation Algorithm}
\label{subsec-to-perm}
For OFA($S,1$), the permutation algorithm (denoted by $\mathcal{P}$) is known as
the best possible algorithm.
For OFAL($S,c$),
let $\sigma=r_1\ldots r_n$ be a request sequence
and $M_i$ be an optimal matching of $r_1,\ldots ,r_i$ for $i=1,\ldots ,n$.
Define $S_i$ ($i=1,\ldots ,n$) to be the set of servers included in $M_i$
and $S_0$ to be $\emptyset$ for convenience.
It is known that there are a sequence of optimal matchings $\{M_i\}_{i=1}^n$
such that $S_i \setminus S_{i-1}$ is singleton for $i=1,\ldots ,n$
\cite{KalP1993}.

When an $i$-th request $r_i$ is revealed,
$\mathcal{P}$ determines the server $s$ such that
$S_{i}\setminus S_{i-1}=\{s\}$ and matches $r_i$ with $s$.
For the competitive ratio of $\mathcal{A}^*$ and $\mathcal{P}$,
we also have the following theorem, which implies
that there is a server layout
where $\mathcal{A}^*$ performs well
and $\mathcal{P}$ performs poorly.

\begin{theorem} \label{thm-to-perm}
For any $\epsilon>0$,
define $S=\{s_1, \ldots ,s_{2k}\}$ as follows:
for $i=1,\ldots ,k$,
\begin{align*}
s_{k+i} &= \frac{1-\delta^i}{1-\delta}, \\
s_{k-i+1} &=-\frac{1-\delta^i}{1-\delta},
\end{align*}
where $\delta>0$ is taken
to satisfy $\delta^k+\delta(4k-1)<\epsilon$ and
$(1-\delta)^{-1}<1+\epsilon/2$.
For the server layout $S$,
$\mathcal{A}^*$ is $(3+\epsilon)$-competitive and
the competitive ratio of $\mathcal{P}$ is at least $4k-1-\epsilon$.
\end{theorem}
The proof of Theorem \ref{thm-to-perm}
can be found in Appendix \ref{pf-to-perm}.

\section{A Lower Bound on the Competitive Ratio of MPFS}
\label{sec-lower-mpfs}
%
In this section, we derive a tight lower bound on the competitive ratio 
of algorithms in ${\cal MPFS}$.
In other words, we will show that the following theorem.

\begin{theorem} \label{thm-lb-mpfs}
Let $\mathcal{A}\in\mathcal{MPFS}$ for $\mathrm{OFAL}(S,c)$.
Then, $\mathcal{R}(\mathcal{A})\geq 2\alpha(S)+1$, i.e.
for any $\mathcal{A}\in\mathcal{MPFS}$ w.r.t. $\mathrm{OFAL}(S,c)$,
there exists a request sequence $\sigma$ such that
\[
\mathcal{A}(\sigma|S)\geq(2\alpha(S)+1)\mathrm{Opt}(\sigma|S).
\]
\end{theorem}
To prove Theorem \ref{thm-lb-mpfs}, the following lemma \cite{HIM2023}
is useful.
\begin{lemma} \label{lem-lb}
Let $\mathcal{A} \in {\cal MPFS}$ for ${\rm OFAL}(S,c)$. Then, 
there exists a request sequence $\sigma$ such that
$
\mathcal{A}(\sigma|S)\geq(2L(S)+1)\mathrm{Opt}(\sigma|S).
$
\end{lemma}
By the definition of $L(S)$ and $\alpha(S)$ in (\ref{ls-alphas}),
we have $L(S)\leq \max_{T\subseteq S} L(T)=\alpha(S)$.
Therefore, Theorem \ref{thm-lb-mpfs} is the improvement
of Lemma \ref{lem-lb}.
\vskip.5\baselineskip
\noindent
\textbf{Proof of Theorem \ref{thm-lb-mpfs}:}
Fix any $\mathcal{A} \in \mathcal{MPFS}$ for ${\rm OFAL}(S,c)$ arbitrarily.
By Lemma \ref{lemma-surrounding-alg}, it suffices to consider the case where
$\mathcal{A}$ is surrounding-oriented.
Let $S'\subseteq S$ be a set of servers such that
$L(S')=\max_{T\subseteq S}L(T)=\alpha(S)$. 

Define a request sequence $\sigma$ as follows:
for each $s\in S\setminus S'$, we first give $c(s)$ requests on $s$.
Since $\mathcal{A}$ is surrounding-oriented, both $\mathcal{A}$ and Opt
match a request on $s\in S\setminus S'$ with $s$ and are incurred no cost
at this stage. Next, we give a request sequence $\sigma'$ for OFAL($S',c$) which 
satisfies the condition of Lemma \ref{lem-lb}, i.e.
$
\mathcal{A}(\sigma'|S')\geq(2L(S')+1)\mathrm{Opt}(\sigma'|S')
$.
Then, we have
\begin{align*}
\mathcal{A}(\sigma|S) &=\mathcal{A}(\sigma'|S')
\geq(2L(S')+1)\mathrm{Opt}(\sigma'|S')
=(2\alpha(S)+1)\mathrm{Opt}(\sigma|S).
\end{align*}
Since $L(S')=\alpha(S)$ and $\mathrm{Opt}(\sigma'|S')=\mathrm{Opt}(\sigma|S)$,
this completes the proof.
\BQED
\vskip.5 \baselineskip
By Theorem \ref{thm-lb-mpfs},
the PTCP algorithm $\mathcal{A}^*$
turns out to be best possible
among all MPFS algorithms, and thus
we have $\mathcal{R}(\mathcal{A}^*)=2\alpha(S)+1$.

\section{Concluding Remarks and Open Questions} \label{sec-conclusion}
In this paper,
we dealt with the online facility assignment problem
on a line OFAL($S,c$) where $S$ is a set of servers
and $c:S\to\mathbb{N}$ is a capacity of each server.
In Section \ref{sec-def-alg}, we proposed
a new MPFS algorithm PTCP (Policy Transition
at Critical Point) and showed that
for OFAL($S,c$),
PTCP is $(2\alpha(S)+1)$-competitive
(in Theorem \ref{thm-main}),
where $\alpha(S)$ is given in (\ref{ls-alphas}).
In Section \ref{sec-lower-mpfs},
we showed that the competitive ratio of
any MPFS algorithm is at least $2\alpha(S)+1$
(in Theorem \ref{thm-lb-mpfs}), i.e.,
PTCP is the best possible MPFS algorithm
for OFAL($S,c$).

However, it is not known if there is an algorithm
$\mathcal{A}\notin\mathcal{MPFS}$ whose
competitive ratio is less than $2\alpha(S)+1$.
Moreover, we do not even know whether
there exists an algorithm $\mathcal{A}\notin\mathcal{MPFS}$
with the capacity-insensitive property
for OFA($S,c$) or not.
Specifically, it would be interesting to study
whether the competitive ratio of
the permutation algorithm \cite{KalP1993, ARK2020}
or the Robust-Matching algorithm \cite{R2016}
for OFA($S,c$) depends on a capacity $c$
or not.


%
\appendix
\section{Proof of Claim \ref{claim1}} \label{app-pf-claim1}

Since $\sigma$ is opposite, $s_{\mathcal{A}_{d,x}}(r_i;\sigma)\leq r_i$ holds.
Let $s'_1<s'_2<\cdots<s'_p=s^*$ be all the servers
in $F_{i-1}(\mathcal{A}_{d,x})$ that locate between $r_i$ and $s_{k+1}$.
For $j=1,\ldots ,p$, let $\sigma'_j$ be a reqest sequence obtained by moving
the position of $r_i$ to $s'_j$.
Note that $\sigma'_p=\sigma^*$ in Claim \ref{claim1}.
Consider a hybrid algorithm $\mathcal{H}_{i,s'_1}$
of $\mathcal{A}_{d,x}$.
Since $h_i=s_{\mathcal{A}_{d,x}}(r_i;\sigma)\leq r_i<s'_1=a_i$, we have
\[
h_{t^*}\leq\cdots\leq h_i\leq r_i<a_i\leq\cdots\leq a_{t^*} 
\]
by Lemma \ref{lem-line-atht}.
For the case $h_{t-1}\neq h_t$, it follows that
\begin{align*}
|r_t-s_{\mathcal{A}_{d,x}}(r_t;\sigma)|-|r_t-s_{\mathcal{H}_{i,s'_{1}}}(r_t;\sigma)|
&=|r_t-h_{t}|-|r_t-h_{t-1}|\leq h_{t-1}-h_t.
\end{align*}
For the case $a_{t-1}\neq a_t$, we get
\begin{align*}
|r_t-s_{\mathcal{A}_{d,x}}(r_t;\sigma)|-|r_t-s_{\mathcal{H}_{i,s'_{1}}}(r_t;\sigma)|
&=|r_t-a_t|-|r_t-a_{t-1}|= a_{t-1}-a_t
\end{align*}
since $r_i$ is the rightmost request in $\sigma$ and $r_t\leq r_i< a_{t-1}\leq a_{t}$.
If $t=t^*+1$, then
\[
|r_{t^*=1}-s_{\mathcal{A}_{d,x}}(r_{t^*+1};\sigma)|-|r_{t^*+1}-s_{\mathcal{H}_{i,s'_1}}(r_{t^*+1};\sigma)|
=|r_{t^*+1}-a_{t^*}|-|r_{t^*+1}-h_{t^*}| \leq |a_{t^*}-h_{t^*}|.
\]
Thus, we have
\begin{align*}
&\mathcal{A}_{d,x}(\sigma|\tilde{S}) - \mathcal{H}_{i,s'_1}(\sigma|\tilde{S})
= \sum_{t=i}^{t^*+1} \left( |r_t-s_{\mathcal{A}_{d,x}}(r_t;\sigma)|-|r_t-s_{\mathcal{H}_{i,s'_1}}(r_t;\sigma)| \right) \\
&\hspace{2cm}\leq |r_i-h_i|-|r_i-a_i|+ \sum_{t=i+1}^{t^*} \left(a_{t-1}-a_t + h_{t-1}-h_t \right) + |a_{t^*}-h_{t^*}| \\
&\hspace{2cm}= (r_i-h_i)-(a_i-r_i)+(a_i-a_{t^*})+(h_i-h_{t^*}) + (a_{t^*}-h_{t^*}) \\
&\hspace{2cm}= 2\,(r_i-h_{t^*}) \leq 2\,\alpha(S)|r_i-a_i|.
\end{align*}
The last inequality is due to the fact that $\mathcal{A}$ satisfies (C3)
in Definition \ref{def-c1c2}.
Hence, it follows that
\begin{align}
\mathcal{A}_{d,x}(\sigma|\tilde{S}) - \mathcal{A}_{d,x}(\sigma'_1|\tilde{S})
&=\mathcal{A}_{d,x}(\sigma|\tilde{S}) -
(\mathcal{H}_{i,s'_1}(\sigma|\tilde{S})-|r_i-a_i|) \nonumber \\
&\leq (2\alpha(S)+1)|r_i-a_i| \nonumber \\
&= (2\alpha(S)+1)|r_i-s'_1|. \label{app1}
\end{align}
Similarly to the above discussion, 
\begin{align}
\mathcal{A}_{d,x}(\sigma'_j|\tilde{S}) - \mathcal{A}_{d,x}(\sigma'_{j+1}|\tilde{S})
\leq (2\alpha(S)+1)|s'_j-s'_{j+1}|
\label{app2}
\end{align}
holds for $j=1,\ldots ,p-1$.
By summing up the formulas (\ref{app2}) for $j=1,\ldots ,p-1$ and (\ref{app1}),
we finally obtain
\[
\mathcal{A}_{d,x}(\sigma|\tilde{S}) - \mathcal{A}_{d,x}(\sigma^*|\tilde{S}) \leq (2\alpha(S)+1)|r_i-s^*|.
\]
%
\section{Deferred Proofs in Subsection \ref{subsec-comparison}}
\label{pf-comparison}
\subsection{Proof of Theorem \ref{thm-to-greedy}}
\label{pf-to-greedy}
Obviously, $\alpha(S)$ is $2$. Then, we have that $\mathcal{A}^*$ is
$2\alpha(S)+1=5$-competitive.

We next show that for any $\epsilon>0$,
there exists a request sequence $\sigma$ such that
\[
\frac{\mathcal{G}(\sigma|S)}{\mathrm{Opt}(\sigma|S)} \geq 2^k-1-\epsilon.
\]
Take $\delta>0$ such that $k\delta/(1+k\delta)\leq \epsilon\cdot 2^{-k}$.
Define $\sigma$ as follows:
for each $s\in S$,
we first give $c(s)-1$ requests on $s$
and next give $k$ requests $r_1, \ldots ,r_k$ where $r_i=2^{i-1}+\delta$ for $i=1,\ldots ,k$.
Then, 
$\mathcal{G}$ matches $r_i$ with $s_{i+1}$
for $i=1,\ldots ,k-1$ and $r_k$ with $s_1$
by the definition of $\mathcal{G}$.
Hence, we have
\[
\mathcal{G}(\sigma|S) \geq 1+2+\cdots +2^{k-1}-k\delta = 2^k-1-k\delta
\]
and $\mathrm{Opt}(\sigma|S)\leq1+k\delta$.
Then, it follows that
\begin{align*}
\frac{\mathcal{G}(\sigma|S)}{\mathrm{Opt}(\sigma|S)}
\geq \frac{2^k-1-k\delta}{1+k\delta} 
= 2^k-1-\frac{k\delta\cdot 2^k}{1+k\delta}
\geq 2^k-1-\epsilon.
\end{align*}
This complete the proof of the theorem. 
%
%
%
%
%
%
\subsection{Proof of Theorem \ref{thm-to-perm}}
\label{pf-to-perm}

It is immediate to see that
\[
2\alpha(S)+1\leq 2\cdot\frac{1}{1-\delta}+1 \leq 2\cdot\left(1+\frac{\epsilon}{2}\right)+1 = 3+\epsilon
\]
and then, $\mathcal{A}^*$ is $(3+\epsilon)$-competitive.

We next show that there exists a request sequence $\sigma$
such that
$\mathcal{P}(\sigma|S)\geq (4k-1-\epsilon)\mathrm{Opt}(\sigma|S)$.
For each $s\in S$, our input $\sigma$ gives $c(s)-1$ requests on $s$
and next gives $2k$ requests $r_1,\ldots ,r_{2k}$.
For $i=1,\ldots ,k$,
define $r_{2i-1}$ to be $x_{2i-1}-\epsilon_{2i-1}$
and $r_{2i}$ to be $x_{2i}+\epsilon_{2i}$, where
\begin{align*}
x_{2i-1}&=\frac{s_{k+i-1}+s_{k+i}}{2}, \\
x_{2i}&=\frac{s_{k-i}+s_{k-i+1}}{2},
\end{align*}
and for $j=1,\ldots ,2k$, let
\[
\epsilon_{j}=\frac{1}{2^{2k-j+1}}\cdot\frac{\delta^k}{1-\delta}.
\]
By the definition of $\mathcal{P}$,
it is immediate that $\mathcal{P}$ matches $r_{2i-1}$ with $s_{k-i+1}$
and $r_{2i}$ with $s_{k+i}$ for $i=1,\ldots ,k$.
Thus, we have
\begin{align*}
\mathcal{P}(\sigma|S) &\geq |r_1-s_k|+(2k-1)|s_{k+1}-s_k| \\
&= 1-\epsilon_1+(2k-1)\cdot 2 \\
&= 4k-1-\epsilon_1 \\
&\geq 4k-1-\delta^k(1-\delta)^{-1} \\
\mathrm{Opt}(\sigma|S) &\leq \frac{s_{2k}-s_1}{2} +\epsilon_1+\cdots +\epsilon_{2k} \\
&= \frac{1-\delta^k}{1-\delta} + \frac{\delta^k}{1-\delta}
\left(\frac12 + \cdots + \frac{1}{2^{2k}}\right) \\
&\leq \frac{1}{1-\delta}.
\end{align*}
Therefore, it follows that
\begin{align*}
\frac{\mathcal{P}(\sigma|S)}{\mathrm{Opt}(\sigma|S)}
&\geq \left(4k-1-\delta^k(1-\delta)^{-1}\right)(1-\delta) \\
&= 4k-1-\delta(4k-1)-\delta^k \\
&\geq 4k-1-\epsilon.
\end{align*}
\end{document}